\documentclass[]{interact}

\usepackage{color}
\usepackage{tabularx}
\definecolor{warninggreen}{rgb}{0.3, 0.73, 0.09}

\usepackage{epstopdf}
\usepackage[caption=false]{subfig}

\usepackage[numbers,sort&compress]{natbib}
\bibpunct[, ]{[}{]}{,}{n}{,}{,}
\renewcommand\bibfont{\fontsize{10}{12}\selectfont}
\makeatletter
\def\NAT@def@citea{\def\@citea{\NAT@separator}}
\makeatother

\theoremstyle{plain}

\theoremstyle{definition}

\theoremstyle{remark}

\begin{document}


\title{Macros to Conduct Tests of Multimodality in SAS}
\author{
\name{Zachariah Neville\textsuperscript{a} and Naomi Brownstein\textsuperscript{b}\thanks{CONTACT Naomi Brownstein. Email: naomi.brownstein@med.fsu.edu. Address: Department of Behavioral Sciences \& Social Medicine, College of Medicine, 1115 West Call Street, Tallahassee, FL 32306-4300, USA}}
\affil{\textsuperscript{a}Department of Statistics, Florida State University, Tallahassee, FL, USA; \textsuperscript{b}Department of Behavioral Sciences \& Social Medicine, Florida State University, Tallahassee, FL, USA}
}

\maketitle

\begin{abstract}
The Dip Test of Unimodality and Silverman's Critical Bandwidth Test are two popular tests to determine if an unknown density contains more than one mode. While the tests can be easily run in R, they are not included in SAS software. We provide implementations of the Dip Test and Silverman Test as macros in the SAS software, capitalising on the capability of SAS to execute R code internally. Descriptions of the macro parameters, installation steps, and sample macro calls are provided, along with an appendix for troubleshooting. We illustrate the use of the macros on data simulated from one or more Gaussian distributions as well as on the famous {\it iris} dataset.

Word Count: 7561
\end{abstract}


\section[Introduction]{Introduction}
Determining the number of modes in an unknown density has applications to many fields, including biology \citep{multmodex1}, political science \citep{multmodex2}, and psychology \citep{multmodex3}. The question of whether a data set contains more than one mode is asked frequently in  
clustering \citep{LuHayesNobelMarron,dip_means,Ahmed2012,sigtest}. The presence of multiple modes may indicate that the data was generated from a population with multiple distinct subgroups.

Two popular methods for testing for multimodality include the Dip test by Hartigan and Hartigan \cite{hartigan1985dip} and the Critical Bandwidth test by Silverman \cite{silverman1981using}. Implementations of these tests exist in R, but SAS does not provide procedures for multimodality tests. 

In this paper, we provide implementations of the Dip Test and Silverman Test that utilise the capability for SAS software to execute R statements in \texttt{PROC IML} \citep{SAS-STAT}. 
The two SAS 
macros, \%dip and \%silverman, offer the same functionality as the R implementations, plus 
additional options, such as storing the output in a SAS data set or setting the random number generator seed for the Silverman test. Additionally, the SAS macros perform error checking and setup in SAS before executing R statements to calculate the test statistic and $p$~value, and displaying the results in the Results Viewer. 

First, we present background on the Dip and Silverman tests and discuss current implementations of the tests in the R language. We then explain the details of the \%dip and \%silverman macros in SAS, including steps for configuring the SAS environment, the implementation details of the macros, the parameters available to customise the tests, and the format of the output. Next, we run the macros on some data sets and discuss the results. We close with a conclusion 
discussing limitations of our work and areas for future exploration. Finally, we provide an appendix to help with troubleshooting.

\section{Background}\label{intro}
When examining data from an unknown distribution, users are often interested in whether that underlying distribution is unimodal or multimodal. Many methods are available to answer this question, including the Dip test \cite{hartigan1985dip}, Critical Bandwidth test \cite{silverman1981using}, and the Excess Mass test \cite{muller1991excess}. Minnotte provides an overview of these and other tests \cite{minnotte1993test}. Our efforts focus on implementing the Dip and Silverman tests in SAS. We begin by providing some background on these two tests.

The Dip test computes the maximum difference between the empirical distribution function and the unimodal distribution function that minimises the maximum difference \citep{hartigan1985dip}. This maximum difference is called the \textit{dip}. To test for statistical significance, the dip statistic D is compared to critical values obtained by simulating with uniform distributions. Hartigan and Hartigan show that the distribution of the dip is asymptotically larger for the uniform than any other distribution with exponentially decreasing tails \cite{hartigan1985dip}. The choice of the uniform distribution as a reference yields a conservative test. In fact, in moderate to large samples, the Silverman test has greater power than the the Dip test \citep{cheng1998calibrating}. Calibration of the Dip test was performed by Cheng and Hall \cite{cheng1998calibrating}, but this calibrated version of the test is not available in R, and hence is not available in the \%dip SAS macro. The Dip test is designed for univariate data. An extension of the Dip test for multivariate data was discussed in \cite{hartigan1985dip}, but the extension is not implemented in the R package or in our SAS macro.

Unlike the Dip test, which uses the empirical distribution function of the data, the Silverman critical bandwidth test uses kernel density estimates to obtain a test statistic \citep{silverman1981using}.  The normal density function is used as the kernel for both theoretical and computational reasons \citep{silverman1981using}. The test statistic is the critical window width ($h_{crit}$) defined as the smallest window such that the kernel density has \texttt{k} or fewer modes. Larger values of $h_{crit}$ indicate that more smoothing of the data is necessary to obtain a density with \texttt{k} or fewer modes, thereby suggesting rejection of the null hypothesis. To assess the statistical significance using a $p$~value, bootstrap samples are generated, and the kernel density estimate and window width $h$ are calculated for each sample. The unadjusted test as originally given in Silverman \cite{silverman1981using} is vulnerable to identifying spurious modes in the tails of distributions; this tendency is discussed further in Hall and York \cite{hall2001calibration}. In other cases, the unadjusted Silverman test is conservative and thus can suffer from low power \citep{hall2001calibration}. For the \texttt{k = 1} (unimodal null hypothesis and a multimodal alternative) case, our SAS implementation and the \texttt{silvermantest} \citep{silvermanpkg} package in R use the calibration performed by Hall and York that is more asymptotically accurate. 
The Silverman test is designed only for univariate data.

Additionally, Silverman's test may be used to determine the number of modes in the distribution of the population from which the data was sampled. If the null hypothesis of unimodality is rejected, the user may wish to find out if there are only two modes, or if more modes exist. On the other hand, when the conclusion of the (unadjusted) Silverman test is a failure to reject the unimodal null hypothesis, one should not retry the Silverman test with a larger value of \texttt{k}. This is because a failure to reject for \texttt{k}=1 means there is insufficient evidence to conclude that the data came from a multimodal distribution. As a result, one should not further test the null hypothesis that the data came from a distribution with two or fewer modes, because one would expect the same result as before: a failure to reject. 
In general, the number of modes can be estimated by conducting Silverman tests sequentially, beginning with \texttt{k}=1 and stopping at the first \texttt{k} such that the null hypothesis is not rejected at the {{\it a priori}} specified significance level.

Implementations of both the Silverman test and Dip test are available in R. The Dip test can be performed using the \texttt{diptest} R package, which is currently available for download \citep{diptestpackage}. As of the time of writing of the present paper, the Silverman test is not available for download as an R package. We are aware that the \texttt{silvermantest} package was previously available as an R package through CRAN, but it was only available to download as a package from the original authors' website at the time of writing our macros and documentation \citep{silvermanpkg}\footnote{To create the \%silverman SAS macro, we downloaded the \texttt{silvermantest\_1.0.tar.gz} file and utilised code from the R files inside the package.}.  The uncalibrated test as originally described by Silverman \cite{silverman1981using} is the default selection in both the R package and the SAS macro. For the \texttt{k} = 1 (unimodal vs multimodal) case only, the calibrated version of the test provided by Hall and York \cite{hall2001calibration} is available as an alternative option. The Dip test function in R does not include an option to produce the calibration recommended by Hall and Cheng. 

SAS software does not currently provide implementations of the Dip Test or Silverman Test, nor of any other tests for multimodality. Because neither Dip nor Silverman is currently available in base SAS software, SAS users wishing to perform these tests must transfer their data and run their analysis in a separate R environment before returning to work in SAS. This limitation is problematic, because the user may not be familiar with the R language, it is not convenient to repeatedly export data from SAS and import into R, and it is easier and more reproducible to use only a single program while performing data analyses.

While the \%dip and \%silverman SAS macros require R to be installed on the machine, they allow a SAS user to perform these multimodality tests without leaving the SAS environment. They also work directly with SAS data sets, and the user does not need to transfer data between SAS software and R. The next section provides details on the implementation of our macros.

\section{Macro Implementation}
We implement both \%dip and \%silverman as macros in SAS. The \texttt{macroDefinitions.sas} file contains the macro definitions and must be run before the macros can be used. Once the \texttt{macroDefinitions.sas} file has been successfully executed and the macros initialised, the macros can be used in the SAS session. These SAS macros utilise existing implementations in R, along with the ability of SAS software to execute R statements, in order to add the Dip and Silverman tests to SAS without having to code them from scratch.

\subsection{General System Requirements}
Because both macros call R from the SAS system, 
R statistical software must be installed on either the SAS server (if running in an enterprise environment or using SAS OnDemand for Academics, for example) or on the local machine (if running SAS software on an individual computer), and the \texttt{RLANG} option must be enabled. The \texttt{RLANG} option can only be set at SAS startup; one convenient way is to modify the \texttt{SASV9.cfg} file and insert the \texttt{RLANG} option \citep[p.~241]{SAS-rlang}. Additional information about locating and modifying the configuration file can be found in SAS documentation \citep[p.~12]{SAS-config}. Further details on how to enable the \texttt{RLANG} option are provided in Section \ref{install}. 

The interface to R is supported only on computers running a Windows or Linux operating system. R cannot be called from SAS University Edition \citep[p.~240]{SAS-rlang}. At the time of writing, our initial testing with SAS OnDemand for Academics (SAS ODA) was also unsuccessful; we were unable to modify the \texttt{SASV9.cfg} file and could not execute any R statements on SAS ODA. The SAS Viya software platform allows the execution of code in many languages, including R, and may be capable of running these macros. However, we have not tested this functionality.

The \%dip SAS macro uses the \texttt{diptest} package in R, and the \%silverman SAS macro uses the \texttt{splines} package in R if the user has specified that they wish to use the calibrated version of the Silverman Test. The SAS macros will install the required R packages if they are not already installed on the system. For each macro, either the machine must be connected to the Internet and capable of installing new R packages, or the packages must already be installed on the machine. If the macro needs to install a new package, a pop-up list of CRAN mirrors may appear. After selecting one, the package will automatically download and install. For best results, the user should use the most current release of SAS. Older versions of SAS may have compatibility problems with the \texttt{RLANG} option or the specific version of R installed on the machine.

\subsection{Required Arguments for Both Macros}
The \%dip and \%silverman macros have certain attributes in common. Each has two required arguments: a univariate SAS data set containing only numeric values, and a file path or fileref to the appropriate macro file containing the R statements executed by the macro. All other arguments for each macro are optional.

The data set provided to the \%dip and \%silverman is specified in the \texttt{dipData} and \texttt{silvData} argument, respectively. For both macros, the input must be a one-dimensional SAS data set; the implementations of the Dip and Silverman tests were not meant for multivariate data. If multidimensional data is used in either macro, then that macro generates an error and quits. The data set must be entirely numeric; if non-numeric data is found in the data set, an error is generated in the Log and Results Viewer, and the test is not run. Data sets with certain SAS formats - such as date and time formats - may generate errors when used in the macros. For more details, including steps for removing formats, please see the appendix.

The \texttt{include} argument in the \%dip and \%silverman macros must be a complete file path to the \texttt{dip.sas} or \texttt{silverman.sas} file provided with this paper, respectively.
The \texttt{.sas} extension and quotation marks are required. Using a relative file path, such as \texttt{include = }``\texttt{dip.sas}'', does not work. The \texttt{include} argument is also required, because the \texttt{dip.sas} and \texttt{silverman.sas} files contain all of the \texttt{PROC IML} code (which then contains R code) needed to perform the tests. Missing or improperly specified files result in error messages. Alternatively, a fileref can be associated with the \texttt{dip.sas} or \texttt{silverman.sas} file, and that fileref can be used for the \texttt{include} argument. The \texttt{FILENAME} statement can be used to assign a fileref; more details are available in SAS documentation \citep[p.~149]{SAS-config}. Easy portability is one advantage of using a fileref instead of hard-coding the file path as a string, because the file path needs to be only changed once (where the fileref is assigned) rather than for every single macro call. The \texttt{samples.sas} file provided with this paper uses \texttt{FILENAME} statements. 

\subsection{Installation Steps}\label{install} A summary of steps required to prepare the SAS environment to run the macros is provided in Table \ref{tab:setup}. When running the macros for the first time, the instructions in both sections of Table \ref{tab:setup} must be completed. After that, only the second section needs to be completed in order to use the macros. Additional details for each step are provided below.

\begin{enumerate}
\item \textbf{Ensure that R is installed on the machine.} \\
R software must be installed on the same machine that runs the SAS software. If accessing a SAS workspace server through client software such as SAS Enterprise Guide, then R software must be installed on the SAS server \citep[p.~240]{SAS-rlang}. The interface to R is supported on computers running Windows or Linux operating systems.

\item \textbf{Open the \texttt{SASV9.cfg} file and ensure that the \texttt{RLANG} option has been inserted.} \\
More details on locating and modifying the \texttt{SASV9.cfg} configuration file can be found in the SAS Companion \citep[p.~12]{SAS-config}. One possible location may be: \texttt{C:\textbackslash Program Files\textbackslash SASHome\textbackslash SASFoundation\textbackslash 9.4\textbackslash nls\textbackslash en\textbackslash sasv9.cfg}.
Adding \texttt{-RLANG} to the configuration file will suffice; place it near the top of the file. An excerpt from a sample \texttt{SASV9.cfg} file is shown in Figure \ref{fig:rlang}, with line numbers on the left side showing that \texttt{-RLANG} is placed at the top of the config file. The other configuration options in Figure \ref{fig:rlang} are only shown to provide context. The \texttt{RLANG} option needs to be specified at startup. After adding the \texttt{RLANG} option to \texttt{SASV9.cfg}, the SAS software should be restarted so that the updated configuration is used. 

[Figure \ref{fig:rlang} near here.]

\item \textbf{If using the \%dip macro or the adjusted option of the \%silverman macro: Verify that the machine can install new R packages from the Internet or that the appropriate R packages are installed on the machine.} \\
The Dip test requires the \texttt{diptest} R package to be installed. When using the adjusted Silverman test, the \texttt{splines} R package must be installed. If these packages are not already installed on the computer, the SAS macros will install them automatically. Therefore, the machine must already have the packages installed or the machine must be capable of installing these packages.

\item \textbf{Download the supplemental code files and store \texttt{dip.sas} and \texttt{silverman.sas} in a safe location for future use.} \\
Download the attached supplemental files. The \texttt{dip.sas} and \texttt{silverman.sas} files are necessary to run the \%dip and \%silverman SAS macros, respectively. These files should be stored in an easily accessible folder for future use. Because the \%dip and \%silverman macros need to know the file paths to the \texttt{dip.sas} and \texttt{silverman.sas} files,  respectively, storing the files in a temporary location, such as the Downloads folder, may not be ideal.

\item \textbf{Execute the \texttt{macroDefinitions.sas} file to add the macros to the operating environment.} \\
The \texttt{macroDefinitions.sas} file contains the macro definitions. After the file has been executed, the macros will be available for use in the current SAS session. Users who plan to run these macros regularly may choose to store them for use between sessions by using the autocall macro facility or the stored compiled macro facility. More details are available in the SAS Macro Language: Reference documentation \citep[p.~113]{SAS-macroref}.

\item \textbf{Upon successful completion of the prior steps, the macros can be used as described in the paper.} \\
The \texttt{samples.sas} file contains replication code for the examples shown in Section \ref{examples}.

\end{enumerate}

[Table 1 near here.]

\subsection{Dip Macro}
\label{subsec:dip}
The following is the full macro call with all of the parameters and default values listed:

\texttt{\%dip(dipData =, simulatepvalue = 0, reps = 2000, out = , completecase = 0, include = )}

The \%dip macro takes several user arguments, with two of them required: a data set (\texttt{dipData}) and a file path to the \texttt{dip.sas} file (\texttt{include}). It performs some basic error checking and setup in SAS and then transfers the arguments and data set to R, where it performs the Dip Test of Unimodality. 

A list of the arguments to the \%dip macro and their descriptions is provided in \ref{dip:args} below. For arguments that have a default value, the default is the same in the SAS macro as in the \texttt{diptest} R package. Note that, because SAS software does not have Boolean data types, we use 0 in place of FALSE and 1 in place of TRUE.

The \texttt{completecase} argument allows for the macro to perform a complete case analysis. While the Dip Test implementation (in R, and thus, in our SAS macro) ignores missing values in the data set by default, the \texttt{completecase} argument is added to be consistent with the \%silverman SAS macro and Silverman Test, which require missing data to be removed before running the test. If missing data is present and \texttt{completecase} is either not specified or set equal to zero, then the macro will stop and return an error. If \texttt{completecase} is set to 1, then a complete case analysis will be performed and missing data ignored while performing the Dip test.


The results from running the Dip test in R using the \texttt{diptest} package and from running the Dip test using the \%dip macro are the same. This allows for reproducibility of results for users of both R and SAS. The \%dip macro will output the same data as the Dip test in R: the name of the test, the Dip statistic D, the $p$~value, and the alternative hypothesis are all output to the Results Viewer. Error and warning messages are displayed both in the Results Viewer and the Log. 
Messages output to the SAS Log follow the same color scheme as in base SAS: {\color{red}{red}} for errors, {\color{warninggreen}{green}} for warnings, and {\color{blue}{blue}} for successful execution.

\subsubsection{Arguments for the \%dip macro}\label{dip:args}
Below is a complete list of arguments for the \%dip macro, including whether the argument is required and the default value if there is one. For those SAS macro arguments which also exist in the \texttt{diptest} R package \cite{diptestpackage}, the name of the corresponding R argument is given.

\paragraph*{}
\texttt{\underline{dipData}} (named \texttt{x} in R), Required, Default: \textit{none}

\texttt{dipData} specifies the SAS data set to be used in the analysis. All data in \texttt{dipData} must be numeric.

\paragraph*{}
\texttt{\underline{simulatePvalue}} (named \texttt{simulate.p.value} in R), Optional, Default: 0 (FALSE)

A Boolean (0 for FALSE, 1 for TRUE) value indicating whether to simulate $p$~values by Monte Carlo simulation. 

\paragraph*{}
\texttt{\underline{reps}} (named \texttt{B} in R), Optional, Default: 2000

The number of repetitions to be used in Monte Carlo simulation. If \texttt{simulatePvalue} is 0, then this parameter will be ignored.

\paragraph*{}
\texttt{\underline{out}}, Optional, Default: \textit{none}

Specifies a SAS data set in which to store output from the macro, including the dip statistic D, the alternative hypothesis in words, and the $p$~value. There should not be quotes around the data set name. If left blank, then output will be displayed in the SAS Results Viewer window and not stored in any data set.

\paragraph*{}
\texttt{\underline{completecase}}, Optional, Default: 0 (FALSE)

A Boolean (0 for FALSE, 1 for TRUE) value indicating whether to perform a complete case analysis.

\paragraph*{}
\texttt{\underline{include}}, Required, Default: \textit{none}

The complete file path to \texttt{dip.sas}, given as a string and surrounded by quotes, or a SAS fileref associated with the \texttt{dip.sas} file.

\subsection{Silverman Macro}
The following is the full macro call with all of the parameters and default values listed:

\texttt{\%silverman(silvData, k = 1, M = 999, adjust = 0, digits = 6, setSeed = , showSeed = 0, outSeed = , out = , completecase = 0, include = )}

The \%silverman macro takes two required arguments: a SAS data set containing the data to be tested, and the complete file path to the \texttt{silverman.sas} file. Several optional arguments are available to change how the test is performed and control how output is displayed and stored. The macro performs basic error checking in SAS before transferring the data set and arguments to R, where the Silverman Critical Bandwidth test is performed. If applicable, error and warning messages are displayed both in the Results Viewer and the Log. Results of the test are shown in the Results Viewer. Additional parameters for customising the behavior are available, such as providing a seed for the random number generator (RNG) in R or providing a SAS data set to store results from the test.

A list of arguments and their descriptions for the \%silverman macro can be found in \ref{silv:args} below. For arguments that have a default value, the default is the same in the macro as in the \texttt{silvermantest} package obtained from \cite{silvermanpkg}.

Optional arguments to the Silverman macro allow for customisation. 
The \texttt{completecase} argument allows for the macro to perform a complete case analysis. The original code from the \texttt{silvermantest} R package does not run the test when missing data is present: an error is returned. The \texttt{completecase} argument for the SAS macro was added to handle missing data. If \texttt{completecase} is not set or the value is set to 0, and missing data is present, then the macro stops and returns an error. If \texttt{completecase} is set to 1, a complete case analysis is performed and missing data is ignored while performing the Silverman test.

The \texttt{setSeed} argument allows for the user to set the seed for the RNG in R, potentially allowing for reproducible results between different users and environments. Currently, the results between the Silverman test in R and the \%silverman macro in SAS do not perfectly match even if the same seed is set in both environments. However, results within a given environment (either R or SAS) can be consistently reproduced by setting the seed with \texttt{setSeed} argument in the \%silverman SAS macro or calling \texttt{set.seed()} in R. The \texttt{showSeed} argument outputs the state of the RNG to the Results Viewer, and may be useful to compare between machines during troubleshooting to verify that the RNG state is the same on each machine. The \texttt{outSeed} argument outputs the state of the RNG to a specified SAS data set and may prove useful during troubleshooting. 


The \%silverman SAS macro outputs the name of the test, the null hypothesis, and the $p$~value to the Results Viewer. Additional informational messages are output when optional arguments are specified. For example, a message is output if the R seed is set or if the calibrated version of the Silverman test is used.

\subsubsection{Arguments for the \%silverman macro}\label{silv:args}
A complete list of arguments for the \%silverman macro is given below. Whether the argument is required and the default value (if applicable) are also provided. For those SAS macro arguments which also exist in the \texttt{silvermantest} R package \cite{silvermanpkg}, the name of the corresponding R argument is given.

\paragraph*{}
\texttt{\underline{silvData}} (named \texttt{x} in R), Required, Default: \textit{none}

The SAS data set to be used in the Silverman test. Must be one-dimensional and numeric.

\paragraph*{}
\texttt{\underline{k}} (named \texttt{k} in R), Optional, Default: 1

The number of modes to be tested. Null hypothesis $H_{0}$: number of modes $\leq k$. Must be an integer value.

\paragraph*{}
\texttt{\underline{M}} (named \texttt{M} in R), Optional, Default: 999

The number of bootstrap replications. An integer value.

\paragraph*{}
\texttt{\underline{adjust}} (named \texttt{adjust} in R), Optional, Default: 0 (FALSE)

A Boolean value indicating whether to adjust $p$~values using the work by Hall and York \cite{hall2001calibration}. Use \texttt{adjust} = 0 for FALSE (do not adjust) and 1 for TRUE (adjust). The adjustment is performed only when \texttt{k} = 1.

\paragraph*{}
\texttt{\underline{digits}} (named \texttt{digits} in R), Optional, Default: 6

Number of digits for rounding the $p$~value. Only applicable when \texttt{adjust} = 1.

\paragraph*{}
\texttt{\underline{setSeed}}, Optional, Default: \textit{none}

A number to be passed as an argument to the \texttt{set.seed()} function in R. This sets the seed of the random number generator (RNG) in R.
 
\paragraph*{}
\texttt{\underline{showSeed}}, Optional, Default: 0 (FALSE)

A Boolean value (0 for FALSE, 1 for TRUE) indicating whether to show the result of \texttt{.Random.seed} from R (returns a vector with the current random number generator state). Used primarily for troubleshooting.

\paragraph*{}
\texttt{\underline{outSeed}}, Optional, Default: \textit{none}

The name of a SAS data set to store the result of \texttt{.Random.seed} from R. Used primarily for troubleshooting.

\paragraph*{}
\texttt{\underline{out}}, Optional, Default: \textit{none}

Specifies a SAS data set that will contain output from the macro, including the null hypothesis, the number of modes tested, and the $p$~value. If left blank, then output will be displayed in the SAS Results Viewer window and not stored in any data set.

\paragraph*{}
\texttt{\underline{completecase}}, Optional, Default: 0 (FALSE)

A Boolean value (0 for FALSE, 1 for TRUE) indicating whether to perform a complete case analysis.

\paragraph*{}
\texttt{\underline{include}}, Required, Default: \textit{none}

The complete file path to \texttt{silverman.sas}, given as a string and surrounded by quotes, or a SAS fileref associated with the \texttt{silverman.sas} file.

\section{Examples}\label{examples}
We demonstrate the use of the \%dip and \%silverman macros on a variety of data sets. Section \ref{normalexamples} focuses on data simulated from one or more Gaussian distributions. Section \ref{irisexample} highlights the performance of the macros on a famous dataset used in computer science containing measurements from iris flowers described in \cite{fisher1936use}. Within both of these subsections, we describe the data sets, provide the code and output to conduct the \%dip and \%silverman macros on the data sets, and summarise the results.

\subsection{Simulated Gaussian 
Data}\label{normalexamples}
The first example includes a series of data sets with 300 observations generated from a single standard normal distribution or small mixture of univariate normal distributions with unit variance. Four data sets are discussed: one with a single normal distribution (\textit{oneNorm}), two with two normal distributions (\textit{twoNorms1} and \textit{twoNorms2}), and one with three normal distributions (\textit{threeNorms}). 
Data sets consisting of multiple normal distributions contain equal proportions of data from each component distribution (150 observations from each distribution for the mixture of two normal distributions, and 100 observations from each distribution for the mixture of three normals). 
The components of the mixtures in \textit{twoNorms1} are centered at 0 and 2. Similarly, the components of \textit{twoNorms2} are centered around 0 and 4. The final example, \textit{threeNorms}, consists of a mixture of three normal distributions with means of 0, 3.5, and 7. 

The SAS code to generate these data sets is provided in the \texttt{samples.sas} file; the random number generator is set for each data set to facilitate reproducibility. Before the user can perform any of the tests, the user should 
open and execute the \texttt{macroDefinitions.sas} file. Execution creates both macros and makes them available for use, as detailed in Section \ref{install} and Table \ref{tab:setup}. Then, to replicate the examples in this section, the user may open the \texttt{samples.sas} file. The first portion of the file is dedicated to creating the data sets used in these examples, and the remainder calls the macros. 

\subsubsection{Dip Macro with Simulated Gaussian Data}
The first goal of the analysis is to use the Dip test to test each of these four SAS data sets for multimodality. The macro calls shown below achieve this goal:

\texttt{\%dip(oneNorm, include = "C:\textbackslash Documents\textbackslash SAS\textbackslash dip.sas")}

\texttt{\%dip(twoNorms1, include = "C:\textbackslash Documents\textbackslash SAS\textbackslash dip.sas")}

\texttt{\%dip(twoNorms2, include = "C:\textbackslash Documents\textbackslash SAS\textbackslash dip.sas")}

\texttt{\%dip(threeNorms, include = "C:\textbackslash Documents\textbackslash  SAS\textbackslash dip.sas")}
\paragraph*{}
The file path given in the \texttt{include} argument needs to be updated to reflect the location of the \texttt{dip.sas} file on the user's specific machine or server. In this example, the optional arguments \texttt{simulatepvalue}, \texttt{reps}, and \texttt{out} are not specified. To further customise the behavior of the Dip test and output from the test, the user can modify these parameters as desired. Descriptions of these parameters are given in \ref{dip:args}.

\subsubsection{Silverman Macro with Simulated Gaussian Data}
The second goal of the analysis is to use Silverman's test to determine, for each of these four SAS data sets, whether the data was generated from a multimodal distribution. 
The first round of tests uses the default value of \texttt{k} = 1, which tests whether the data came from a unimodal or multimodal distribution. Because the Hall and York calibration \cite{hall2001calibration} is available for the \texttt{k} = 1 case, we also present results for the adjusted Silverman test.
The macro calls to test the null hypothesis of unimodality are shown below. The first four use the unadjusted version of the Silverman test, while the next four use the adjustment. Including both sets of calls illustrates the effect of the adjustment on the results. 

The \texttt{setSeed} argument is provided for all of the calls to the \%silverman macro in these examples to facilitate reproducibility of the results. 

\texttt{\%silverman(oneNorm, setSeed = 1234, include = "C:\textbackslash Documents\textbackslash SAS\textbackslash silverman.sas")}

\texttt{\%silverman(twoNorms1, setSeed = 1234, include = "C:\textbackslash Documents\textbackslash SAS\textbackslash silverman.sas")}

\texttt{\%silverman(twoNorms2, setSeed = 1234, include = "C:\textbackslash Documents\textbackslash SAS\textbackslash silverman.sas")}

\texttt{\%silverman(threeNorms, setSeed = 1234, include = "C:\textbackslash Documents\textbackslash SAS\textbackslash silverman.sas")}

\texttt{\%silverman(oneNorm, setSeed = 1234, adjust = 1, include = "C:\textbackslash Documents\textbackslash SAS\textbackslash silverman.sas")}

\texttt{\%silverman(twoNorms1, setSeed = 1234, adjust = 1, include = "C:\textbackslash Documents\textbackslash SAS\textbackslash silverman.sas")}

\texttt{\%silverman(twoNorms2, setSeed = 1234, adjust = 1, include = "C:\textbackslash Documents\textbackslash SAS\textbackslash silverman.sas")}

\texttt{\%silverman(threeNorms, setSeed = 1234, adjust = 1, include = "C:\textbackslash Documents\textbackslash SAS\textbackslash silverman.sas")}

\paragraph*{}
The third and final goal is to determine the number of modes in the distribution for each of these four SAS data sets. 
The application of Silverman's test with larger values of \texttt{k} is shown for data sets for which the null hypothesis of unimodality was rejected. The adjustment by Hall and York \cite{hall2001calibration} is only available for the \texttt{k} = 1 case; thus, the remaining tests in this example are restricted to the unadjusted version. The macro calls are included below:

\texttt{\%silverman(twoNorms2, k = 2, setSeed = 1234, include = "C:\textbackslash Documents\textbackslash SAS\textbackslash silverman.sas")}

\texttt{\%silverman(threeNorms, k = 2, setSeed = 1234, include = "C:\textbackslash Documents\textbackslash SAS\textbackslash silverman.sas")}


\texttt{\%silverman(threeNorms, k = 3, setSeed = 1234, include = "C:\textbackslash Documents\textbackslash SAS\textbackslash silverman.sas")}

\subsubsection{Results for Simulated Gaussian Data}

[Figure \ref{fig:dip} near here.]

The output from the first \%dip macro call is displayed in Figure \ref{fig:dip}. The results from the \%dip and \%silverman macro calls on the simulated data are summarised in Table \ref{ex:ex1}. The Dip test correctly fails to reject the null hypothesis for the data set \textit{oneNorm}, concluding there is insufficient evidence that the data was generated from a multimodal distribution. For \textit{twoNorms2}, the Dip Test easily rejects the null hypothesis of unimodality, with a $p$~value of {$2.032\times10^{-6}$}. It is very unlikely that data generated from a unimodal distribution would display such a large dip.
On the other hand, for \textit{twoNorms1}, the $p$~value associated with this test is 0.3146, indicating a 
failure to reject the null hypothesis of unimodality. In this case, the means of these two normal distributions are separated by only two standard deviations. Thus, as is evident in a histogram of the data, the modes approach each other and are somewhat more difficult to distinguish visually and statistically.

[Figure \ref{fig:silv} near here.]

[Table 2 near here.]

The output from the first call to \%silverman is displayed in Figure \ref{fig:silv}, and the $p$~values from each of these tests are provided in Table \ref{ex:ex1}. Like the Dip, Silverman's test correctly fails to reject the unimodal null hypothesis for the \textit{oneNorm} data set and rejects the null for the \textit{twoNorms2} data set. 
In contrast to the Dip, while the unadjusted Silverman test fails to reject for the \textit{twoNorms1} data set, the adjusted Silverman test is very close to the 0.05 level, indicating that adjusted Silverman test can more easily distinguish the nearby but distinct modes. 
 In the final data set, \textit{threeNorms}, the difference between the adjusted and unadjusted Silverman tests remains apparent. The unadjusted test is borderline, with a $p$~value of 0.0591, while the adjusted test rejects handily at the 0.05 level, with a $p$~value of 0.0118. The unadjusted Silverman test yields higher $p$~values than the adjusted version, supporting the known finding, discussed in Section \ref{intro}, that the unadjusted Silverman test is conservative \citep{hall2001calibration}.

Results for Silverman's test with \texttt{k} $> 1$ are included in the bottom part of Table \ref{ex:ex1}. This example shows how to use the \%silverman macro not only to classify the modality of the distribution, but also 
to estimate the number of modes of the original distribution. 
The first row displays the test of the null hypothesis that the original distribution (from which \textit{twoNorms2} was sampled) has two or fewer modes against the alternative of three or more modes. The $p$~value of 0.1882 corresponds to correctly failing to reject the null and concluding that the distribution has two or fewer modes. Recall that we obtained a $p$~value of 0 when we did the Silverman test with \textit{twoNorms2} and \texttt{k}~=~1 (see Table \ref{ex:ex1}) which made us suspect the distribution was multimodal. Combining these two pieces of information, we would argue the underlying distribution is likely bimodal, which we know to be correct. 
Similarly, for the \textit{threeNorms} dataset, 
we reject the null hypothesis for \texttt{k}~=~2. Yet, with a $p$~value of 0.6607, we fail to reject the null hypothesis for \texttt{k}~=~3. Therefore, we correctly suspect that the original distribution for \textit{threeNorms} is trimodal.

\subsection{Application to Iris Data}\label{irisexample}
The next example demonstrates the use of the SAS macros on the {\it iris} data set \citep{fisher1936use}, which is found in the \texttt{sashelp} library in the base SAS software and commonly used to demonstrate clustering tasks. The {\it iris} data set includes 150 flowers, each with 4 measurements: sepal length, sepal width, petal length, and petal width. Because the multimodality tests in both SAS and R are only designed for univariate data, we choose to examine the Petal Width. 
It is necessary to construct a new univariate data set containing only the Petal Width data. 
The \texttt{samples.sas} file, included with this paper, provides code to create this data set, denoted {\it irisPW}.

\subsubsection{Dip Macro with Iris Data}
The following call to the \%dip macro executes the Dip test for unimodality on {\it irisPW}:

\texttt{\%dip(irisPW, include = "C:\textbackslash Documents\textbackslash SAS\textbackslash dip.sas")}

\subsubsection{Silverman Macro with Iris Data}
Calls to the \%silverman macro for {\it irisPW} use the code shown below. The first two calls test whether {\it irisPW} was generated from a unimodal distribution. Both the unadjusted and adjusted versions of the Silverman Test are provided to 
illustrate how the results may differ. Additionally, the test is repeated for \texttt{k} $> 1$ to estimate the number of modes in the dataset.

\texttt{\%silverman(irisPW, setSeed = 1234, include = \\ "C:\textbackslash Documents\textbackslash SAS\textbackslash silverman.sas")}

\texttt{\%silverman(irisPW, adjust = 1, setSeed = 1234, include = "C:\textbackslash Documents\textbackslash SAS\textbackslash silverman.sas")}

\texttt{\%silverman(irisPW, k = 2, setSeed = 1234, include = "C:\textbackslash Documents\textbackslash SAS\textbackslash silverman.sas")}

\subsubsection{Results from Iris Data}
Results from the macro calls on {\it irisPW} are shown in Table \ref{ex:ex2}. 
The $p$~value associated with the Dip test is below $2.2\times10^{-16}$, which is extremely low. Table \ref{ex:ex2} also includes very low $p$~values for both the unadjusted and adjusted versions of the Silverman Test for \texttt{k}~=~1. The null hypothesis that the Petal Widths came from a unimodal distribution is rejected for all three tests. The conclusion that the Petal Width distribution contains multiple modes is consistent with the fact that {\it iris} is known to consist of multiple clusters, corresponding to three different species of iris flowers \citep{fisher1936use}. 

[Table 3 near here.]

Additional calls to the Silverman Test facilitate estimation of the number of modes in {\it irisPW}. The use of \texttt{k} = 2 corresponds to a test of the null hypothesis that {\it irisPW} came from a distribution with two or fewer modes, against the alternative that it came from a distribution with three or more modes. 
The second row in Table \ref{ex:ex2} displays the results for \texttt{k} = 2. 
The only available option, the unadjusted Silverman test, yields a 
$p$~value of 0.4775. We fail to reject the null hypothesis that the data has two or fewer modes. 
Consequently, we suspect that the Petal Width distribution is bimodal, not trimodal. Petal Width does vary by species. Indeed, it is smallest for {\it setosa} flowers compared to {\it versicolor} and {\it virginica}. However, the distance between the modes corresponding to {\it setosa} and {\it versicolor} species was much larger than the distance between the modes corresponding to the {\it versicolor} and {\it virginica} species \citep{fisher1936use}. Thus, Silverman's test found insufficient evidence to reject the null hypothesis that {\it irisPW} was generated from a distribution with two or fewer modes, likely because the distributions for the {\it versicolor} and {\it virginica} species were close together and difficult to distinguish.

\section{Concluding Remarks}
The present paper describes two macros, \%dip and \%silverman, to perform the Dip Test and Silverman Test, each of which tests the null hypothesis that a data set was generated from a unimodal distribution. We 
present the structure of the macros, include detailed instructions for how to run the macros and modify parameters to customise results, and display several examples of the macros in action. We also show how the \%silverman macro can help estimate the number of modes in the distribution. Examples, including Gaussian mixtures and the {\it iris} Petal Width measurements, illustrate the capability of and differences between the two macros. Along with the example file (\texttt{samples.sas}), both macros are available and simple for users to download, install, and execute entirely within the SAS environment. Helpful tips for troubleshooting are provided in the appendix.

While our work adds new functionality for users of SAS software, it does have some limitations. One limitation of our work is its dependence on the \texttt{diptest} package and need to call R statements, thus requiring R to be installed on the machine performing the test. The adjusted version of the Dip test is also not included as an option in the SAS macro, because it is not available in the \texttt{diptest} package. In addition, the \texttt{RLANG} option cannot be readily changed in SAS Studio as part of SAS Online for Academics, or in the SAS University Edition software. These limitations make it impossible to run our macros in these popular environments. 

Another potential limitation of our macro is its restriction to univariate data. Both the Dip and Silverman tests in R can technically accept multivariate data (in the form of an R matrix, for example). However, in their implementation, both tests perform data sorting, which merges all of the data into a single long vector. Often, the columns of multivariate data represent distinct variables measured on different scales. In such cases, collapsing the columns results in a single long vector with data whose combination lacks practical meaning. Any test performed on the combined vector yields misleading, even meaningless results. Because of this dangerous tendency, we consider the limitation of our SAS macros to only one-dimensional SAS data sets an improvement over the R packages, which conduct tests on the inappropriately merged data without warning the user that columns were combined. 

In fact, multivariate data can still be used with our macros if the user first reduces the data to a single dimension and then runs the multimodality tests on the reduced data set. For example, principal component analysis is a method to reduce data to a smaller dimension while retaining the original essence of the full data set \cite{jolliffe2002pca}. Alternatively, clustering applications frequently summarise data using the set of distances between each pair of points in the data set. Examples of combining dimension reduction techniques with multimodality tests include the Dip test on the set of pairwise distances \citep{dip_means} and Silverman's test with the first principal component \cite{Ahmed2012}. Additional combinations are being studied in an area of computer science called clusterability, which is an active research area about a critical but rarely used pre-validation step in the clustering process \citep{clusterability}.

Implementation of these methods -- and any new methods that utilise the Dip test or Silverman's test -- in popular programming environments, such as SAS, is an avenue of future research that depends on the macros in the present paper.

We hope that users will find these macros useful for their SAS programming and research needs. Ultimately, however, we hope that the Dip and Silverman tests will be incorporated into base SAS software in the future to facilitate continuous maintenance and compatibility with newer releases of SAS, as well as access for users on all SAS interfaces. 

\bibliographystyle{tfnlm}
\bibliography{myreferences}

\section{Appendix: Troubleshooting}
The following information may be helpful to users when they are beginning to prepare and use the macro. Because we encountered these obstacles in our own testing, we provide the problems and solutions for others.
\begin{itemize}
\item[Q:] I can't edit my SAS config file. What should I do?
\begin{itemize}
\item[A:] Check the permissions on your computer. One option is to edit the config file in a plain text editor, such as Notepad, and run the editor as an administrator. (Right click the editor before opening to allow the option to run as administrator.) You should then be able to save the changes to your config file. 

Alternatively, use the Save As feature of your text editor to save a new copy of the config file, and then navigate to the location of the original and replace it with the new file. 

Make sure to use an ASCII text editor such as Notepad or a SAS text editor to avoid corrupting the config file. Using a specialised editor such as Microsoft Word or WordPad, for example, may corrupt the file \citep[p. 14]{SAS-config}.
\end{itemize}
\item[Q:] I have one or more of the following errors: \\
ERROR: SAS could not initialize the R language interface. \\
ERROR: The installed version of R cannot be used.  The entry point ``ConsoleIob'' could not be located. \\
What can I do?
\begin{itemize}
\item[A:] One cause of these errors is having an out of date version of SAS installed on your computer. (For example, we encountered these errors when running SAS version 9.4 TS1M0, while the most current release at the time of writing was 9.4 TS1M5.)  Please check your Log for your SAS release, and, if needed, upgrade your copy of SAS to the newest release. 
\end{itemize}
\item[Q:] I ran the \texttt{dip.sas} and \texttt{silverman.sas} files and encountered errors. What happened? 
\begin{itemize}
\item[A:]  You do not need to run the \texttt{dip.sas} and \texttt{silverman.sas} files. Doing so will generate errors due to undefined parameters. Running the \texttt{macrodefinitions.sas} file and properly calling the \%dip and \%silverman macros (see the \texttt{samples.sas} file for examples) is sufficient.
\end{itemize}
\item[Q:] I receive the following error when running the \%dip (or \%silverman) macros:

ERROR: File path for include parameter is not valid. Please use the full file path in quotes and include the extension ``.sas''. For example, \\
include = ``C:\textbackslash Documents\textbackslash SAS\textbackslash dip.sas'' 

But I \textit{did} specify the file path. What can I do?
\begin{itemize}
\item[A:] Make sure you have provided the full file path to \texttt{dip.sas} (or \texttt{silverman.sas}, depending on which macro is being used), including the ``.sas'' extension. A relative file path such as ``dip.sas'' will not work. The file path should also be surrounded by double quotes `` ''. 

If you are using a fileref, such as in the \texttt{samples.sas} supplemental file, make sure to follow these same guidelines.
\end{itemize}

\item[Q:]I receive the following error when running the \%dip (or \%silverman) macros:

ERROR: Your data was input in SAS as numeric, but R did not recognise it as numeric. This may be due to the use of SAS formats, such as dates, times, or custom formats. If you decide that removing a format is appropriate, then documentation with directions is available at the following link:
http://support.sas.com/documentation/cdl/en/lrdict/64316/HTML/default/\\viewer.htm\#a000178212.htm

What can I do?

\begin{itemize}
\item[A:] This error is displayed in the Results Viewer when the provided data set is a SAS data set with numeric values, but R did not recognise the data  as numeric. One cause of this error is using a SAS data set with a format applied to one or more of its variables. Certain data formats, such as date and time formats, can cause errors when running the \%dip or \%silverman macros. In this situation, removing the format may resolve the error. Code to remove a format using the \texttt{FORMAT} statement is shown below: 

\begin{verbatim}
data mydata_without_formats;
    set mydataset;
    format x;
run;
\end{verbatim}

where \texttt{mydataset} is the name of your original SAS data set, \texttt{x} is the name of the variable that has the format associated with it, and \texttt{data\_without\_formats} is the name of the newly created unformatted SAS data set. Additional information on SAS formats is available at the link provided in the error message, and in the SAS Language Reference \citep[p.~1579]{sas-languageref}.

\end{itemize}

\end{itemize}

\clearpage
\section{Tables}
\begin{table}[h!]
\begin{tabularx}{\textwidth}{|X|}
\hline
\textbf{Steps to Set Up the \texttt{dip} and \texttt{Silverman} Macros in SAS} \\
\hline
\textbf{First Time} \\
{\textbf{1.}}~~Ensure that R is installed on the machine. \\
{\textbf{2.}}~~Open the \texttt{SASV9.cfg} file and ensure that the \texttt{RLANG} option has been inserted. \\
{\textbf{3.}}~~If using the \%dip macro or the adjusted option of the \%silverman macro: Verify that the machine can install new R packages from the Internet or that the appropriate R packages are installed on the machine. \\
{\textbf{4.}}~~Download the supplemental code files and store \texttt{dip.sas} and \texttt{silverman.sas} in a safe location for future use. \\
\hline
\textbf{Every Time} \\
{\textbf{5.}}~~Execute the \texttt{macroDefinitions.sas} file to add the macros to the operating environment. \\
{\textbf{6.}}~~From here, the macros can be used as described in the paper. \\
\hline
\end{tabularx}
\caption{Instructions for setting up SAS environment to run the macros.\label{tab:setup}}
\end{table}

\clearpage

\begin{table}
\begin{tabularx}{\textwidth}{|X|l|l|l|l|p{57pt}|}
\hline
\textbf{Data Set Name} & \textbf{Description} & \textbf{k} & \textbf{Dip} & \textbf{Silverman} & \textbf{Silverman (adjusted)} \\
\hline
\textit{oneNorm} & One Normal & 1 & 0.7542 & 0.5105 & 0.4351 \\
\hline
\textit{twoNorms1} & Two Normals (Close) & 1 & 0.3146 & 0.1361 & 0.0546 \\
\hline
\textit{twoNorms2} & Two Normals (Far) & 1 & $2.032\times10^{-6}$ & 0$^\dag$ & 0$^*$ \\
\hline
\textit{threeNorms} & Three Normals & 1 & 0.0094 & 0.0591 & 0.0118 \\
\hline
\hline
\textit{twoNorms2} & Two Normals (Far) & 2 & N/A & 0.1882 & N/A \\
\hline
\textit{threeNorms} & Three Normals & 2 & N/A & 0.0010 & N/A \\
\hline
\textit{threeNorms} & Three Normals & 3 & N/A & 0.6607 & N/A \\
\hline
\end{tabularx}
\caption{Summary of output ($p$~values) from running the \%dip and \%silverman SAS macros with various simulations from Normal distributions. \label{ex:ex1}}
$^\dag$ For the unadjusted test, the $p$~value is a bootstrap $p$~value and is not rounded. \\
$^*$ For the adjusted test only, the \%silverman macro rounds any $p$~value below 0.005 down to 0.
\end{table}

\clearpage

\begin{table}
\begin{tabularx}{\textwidth}{|X|p{85pt}|l|l|l|p{60pt}|}
\hline
\textbf{Data Set Name} & \textbf{Description} & \textbf{k} & \textbf{Dip} & \textbf{Silverman} & \textbf{Silverman (adjusted)} \\
\hline
\textit{irisPW} & Petal Width & 1 & $<2.2\times10^{-16}$ & 0.0030$^\dag$ & 0$^*$ \\
\hline 
\textit{irisPW} & Petal Width & 2 & N/A & 0.4775 & N/A \\
\hline
\end{tabularx}
\caption{Summary of results from running the \%silverman and \%dip SAS macros with the \textit{iris} Petal Width feature. \label{ex:ex2}} 
$^\dag$ For the unadjusted test, the $p$~value is a bootstrap $p$~value and is not rounded. \\
$^*$ For the adjusted test only, the \%silverman macro rounds any $p$~value below 0.005 down to 0.
\end{table}

\clearpage
\section{Figures}
\centering
\begin{figure}[h]
\includegraphics[width = \textwidth]{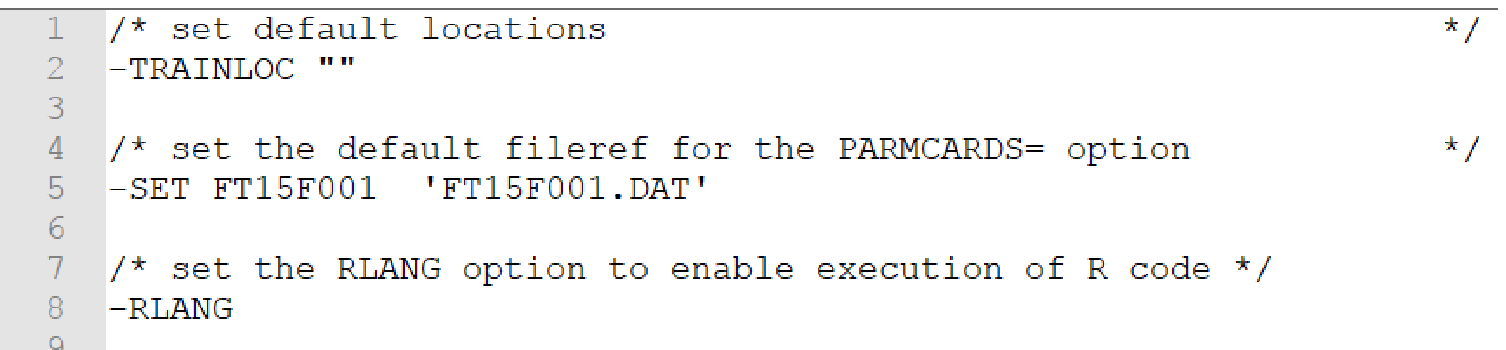}
\caption{The location of the \texttt{RLANG} option in the \texttt{SASV9.cfg} file}
\label{fig:rlang}
\end{figure}

\clearpage
\centering
\begin{figure}[h]
\includegraphics[width = \textwidth]{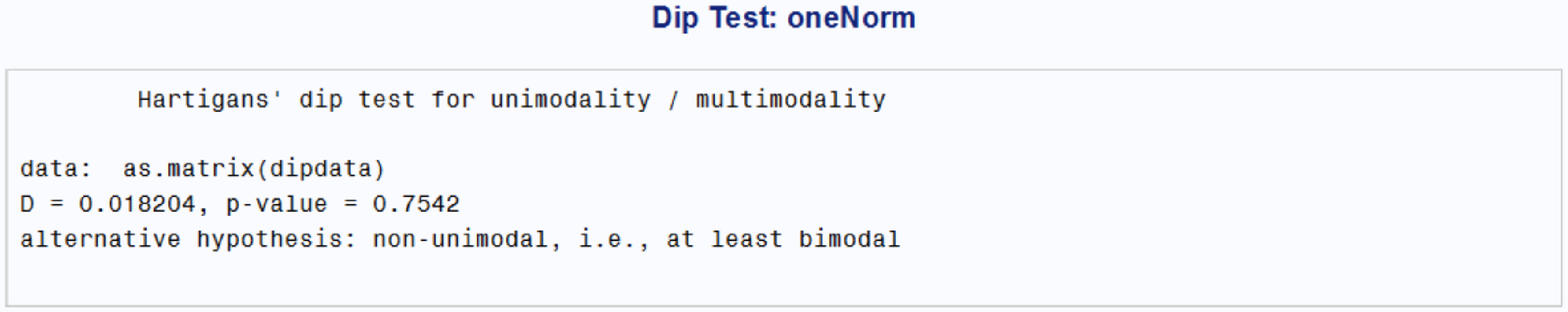}
\caption{Results from the \%dip SAS macro test of unimodality on the \textit{oneNorm} data set.} 
\label{fig:dip}\end{figure}

\clearpage
\centering
\begin{figure}[h]
\includegraphics[width = \textwidth]{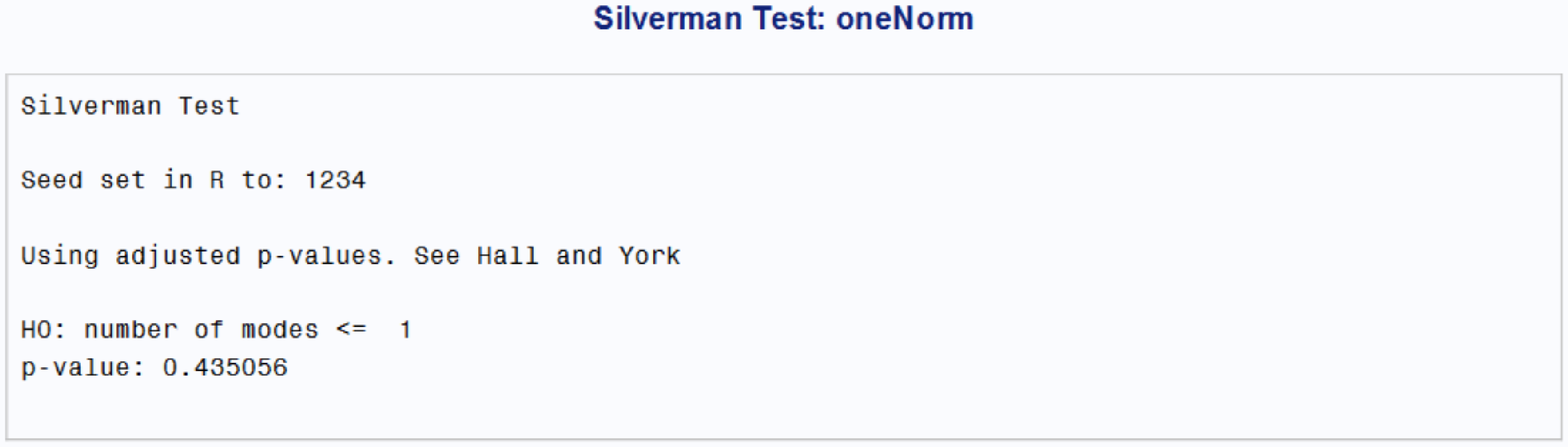}
\caption{Results from the \%silverman macro on \textit{oneNorm}, using the adjusted version from Hall and York \cite{hall2001calibration}.} 
\label{fig:silv}
\end{figure}

\clearpage
\section{Figure Captions}
\paragraph*{}
Figure \ref{fig:rlang}: The location of the \texttt{RLANG} option in the \texttt{SASV9.cfg} file 

\paragraph*{}
Figure \ref{fig:dip}: Results from the \%dip SAS macro test of unimodality on the \textit{oneNorm} data set. 

\paragraph*{}
Figure \ref{fig:silv}: Results from the \%silverman macro on \textit{oneNorm}, using the adjusted version from Hall and York \cite{hall2001calibration}. 
\end{document}